\documentclass[prl,twocolumn,aps,superscriptaddress,showpacs]{revtex4}
\usepackage{graphicx,mathrsfs,bm,times,amsmath}
\makeindex

\begin{document}

\title{Orbital Nematic Instability in the Two-Orbital Hubbard Model: \\ 
Renormalization-Group + Constrained RPA Analysis}

\author{Masahisa Tsuchiizu}
\affiliation{Department of Physics, Nagoya University, Nagoya 464-8602, Japan}

\author{Yusuke Ohno}
\affiliation{Department of Physics, Nagoya University, Nagoya 464-8602, Japan}

\author{Seiichiro Onari}
\affiliation{Department of Applied Physics, Nagoya University, Nagoya 464-8603, Japan}

\author{Hiroshi Kontani}
\affiliation{Department of Physics, Nagoya University, Nagoya 464-8602, Japan}

\date{August 1, 2013}

\begin{abstract}
Motivated by 
the nematic electronic fluid phase
in Sr$_3$Ru$_2$O$_7$, 
we develop a combined scheme of the renormalization-group method and
the random-phase-approximation-type
method, and
analyze orbital susceptibilities of the ($d_{xz},d_{yz}$)-orbital Hubbard model
with high accuracy.
It is confirmed that the present model  exhibits  a
 ferro-orbital
instability near the magnetic or superconducting quantum criticality,
due to the Aslamazov-Larkin-type
vertex corrections.
This mechanism of orbital nematic order presents a natural explanation
for the nematic order in  Sr$_3$Ru$_2$O$_7$,
and is expected to be realized in various multiorbital systems,
such as Fe-based superconductors.
\end{abstract}

\maketitle

A bilayer ruthenate compound Sr$_3$Ru$_2$O$_7$ 
  has attracted much attention since 
it exhibits a unique quantum critical behavior
\cite{Grigera:2001wv,Perry:2001jq,Grigera:2003gl,Mackenzie:2012ct}.
Field-induced antiferromagnetic quantum criticality 
is observed by the many experiments 
such as
the NMR measurements 
\cite{Kitagawa:2005hm}.
Surprisingly, the magnetic quantum critical point is hidden by
  the formation of the nematic electronic liquid phase at very low temperatures
 ($\sim 1$K).
The  orientational-symmetry breaking
in the nematic phase is confirmed by 
the large anisotropy of in-plane resistivity
\cite{Grigera:2004tk,Borzi:2007eg,Mackenzie:2012ct}.
Thus, the quantum criticality and nematic phase
  formation are intimately  linked in this material \cite{Mackenzie:2012ct}.

The band structure of the ruthenate oxides 
are composed of the
Ru $t_{2g}$ ($d_{xy}$, $d_{xz}$, and $d_{yz}$)  orbitals.
For a
microscopic understanding of the nematic phase in 
 Sr$_3$Ru$_2$O$_7$, 
a large number of theoretical works have been devoted.
The spontaneous violation of the $C_4$ symmetry of the Fermi
surface (FS), i.e., Pomeranchuk instability, had been frequently discussed 
 by focusing on the van Hove singularity.
The scenario of 
the single-band Pomeranchuk instability
 was originally proposed by using the renormalization-group (RG) method
\cite{Halboth:2000vm},
and has been analyzed by the mean-field
\cite{Yamase:2007bh,Yamase:2009dx,Fischer:2010fo,Yamase:2010if}
 and perturbation \cite{Yoshioka:2012fz} studies.
However, 
the  temperature-flow ($T$-flow) RG scheme
\cite{Honerkamp:2001vv,Honerkamp:2005fv,Metzner:2012jv}
indicates  that 
the nematic fluctuation
is always weaker than other instabilities.
Thus, the possibility  of the single-band Pomeranchuk instability 
is not settled yet.

An alternative
theoretical route to 
  elucidate the nematic phase is to focus on the two $d_{xz}$ and $d_{yz}$
  orbitals, which 
give rise to 
quasi-one-dimensional
  $\alpha$ and $\beta$ bands.
It has been pointed out that 
the nematic state is 
described as an orbital ordering 
 ($\langle n_{xz}\rangle \neq \langle n_{yz}\rangle$).
This 
ferro-orbital-order
scenario 
has been analyzed 
within the mean-field-level approximation
by focusing only on the $\bm q=\bm 0$ modes
\cite{Raghu:2009bn,Lee:2009fa,Lo:2012vy}.
However, the random phase approximation (RPA)
leads to the occurrence of the antiferro-orbital order 
because of the nesting of the FS \cite{Takimoto:2000ts}.
Therefore, the theoretical analysis beyond the RPA is required.

Recently, 
a
similar nematic phase 
in Fe-based superconductors,
 which are also  multiorbital systems,
has attracted great attention
\cite{Chu:2010cr,Yoshizawa:2012hr}.
Up to now, both the 
spin nematic \cite{Fernandes:2010ci}
and 
orbital nematic  \cite{Onari:2012jb} theories 
had been proposed.
The latter theory pointed out the importance of the 
vertex correction (VC) in the nematic order.
However, 
they studied a 
 limited number of VCs, so
 the importance of VCs should be clarified  by other unbiased 
theoretical techniques.
For this purpose, the RG treatment is quite suitable
because the RG method 
enables
 us to perform the systematic  calculations of 
VCs.

In this Letter,
 we develop an RG scheme and perform 
 accurate evaluations of spin and orbital susceptibilities
in the ($d_{xz}$, $d_{yz}$)-orbital Hubbard model.
We find that  the strong orbital nematic fluctuation, i.e.,
orbital Pomeranchuk instability,
emerges near the magnetic or superconducting quantum criticality
due to the VCs.
The present RG study 
confirms the validity of our previous 
 perturbation analysis in Ref.\ \cite{Ohno:2013hc}.
It is confirmed
that the two-orbital single-layer Hubbard model is a minimal 
  model to describe the orbital nematic order
 realized in Sr$_3$Ru$_2$O$_7$.

We consider a single-layer 
 ($d_{xz}$, $d_{yz}$)-orbital model, where 
the orbital index $\mu=1$ and 2 refer to $d_{xz}$ and $d_{yz}$, respectively.
The tight-binding Hamiltonian with tetragonal symmetry 
is given in the form
\begin{eqnarray}
H_0&=&
\sum_{\bm k,\sigma} \sum_{\mu,\mu'=1,2}
  \xi^{\mu\mu'}_{\bm k} c_{\bm{k},\mu,\sigma}^\dagger c_{\bm{k},\mu',\sigma}^{}
,
\label{eq:H0}
\end{eqnarray}
where $\xi_{\bm{k}}^{11} =-2t\cos k_x -2t_\mathrm{nn} \cos k_y
 + 4 t_\mathrm{nnn} \cos k_x \cos k_y - \bar\mu$,
$\xi_{\bm{k}}^{22} =-2t\cos k_y -2t_\mathrm{nn} \cos k_x
 + 4 t_\mathrm{nnn} \cos k_x \cos k_y - \bar\mu$, and 
$\xi_{\bm{k}}^{12}= \xi_{\bm{k}}^{21}=4t'\sin k_x \sin k_y$,
with $\bar \mu$ the chemical potential.
We set $t=1$ as the energy unit.
We also consider the multiorbital Hubbard interactions
composed of the intra (inter) orbital interaction $U$ ($U'$), and the 
 exchange and pair-hopping interaction $J$.
Throughout the Letter,
the condition  $U=U'+2J$ is assumed.

\begin{figure}[t]
\includegraphics[width=8.5cm]{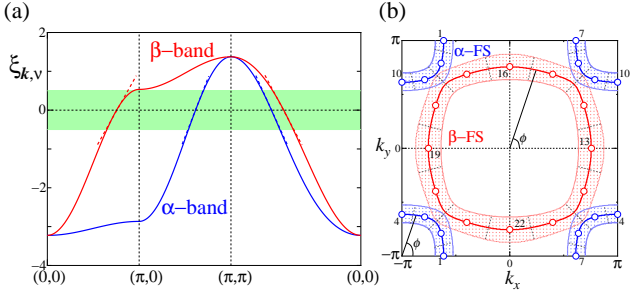}
\caption{
(color online).
(a) Band structure 
of $H_0$ for $n=2.7$.
The linearized band dispersions are shown by the dashed lines.
The low-energy excitations of electrons 
($|\xi_{\bm k,\nu}^\mathrm{linear}| \le \Lambda_0$)
are denoted by the shaded area.
(b) The patch index ($1-24$) on the FSs. 
}
\label{fig:band}
\end{figure}

\begin{figure}[t]
\includegraphics[width=8.5cm]{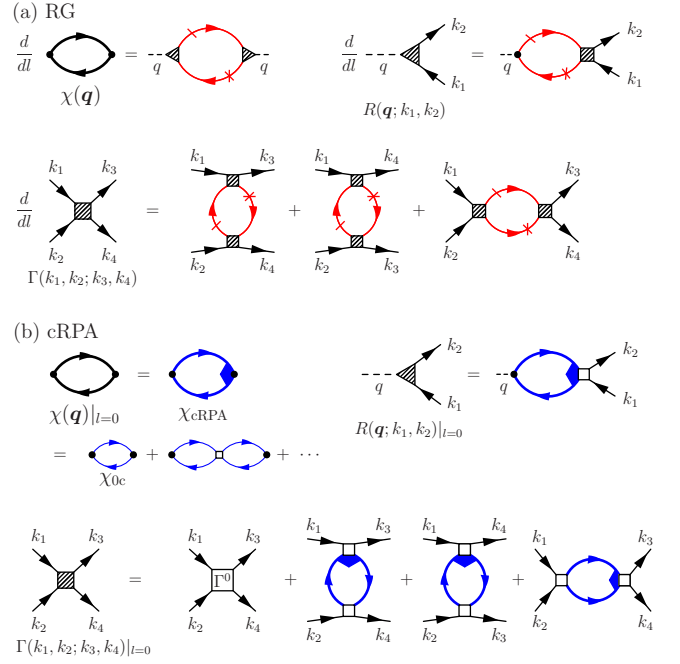}
\caption{
(color online).
The diagrammatic representation of the present  RG+cRPA scheme, where
$l$ is  the scaling parameter.
(a)
The RG equations for the four-point, three-point vertices, and 
the susceptibility.
The slashed (crossed) line represents 
 an electron propagation having the energy 
  on  $\Lambda_{l+dl}<|\xi_{\bm k,\nu}^\mathrm{linear}|<\Lambda_l$
($|\xi_{\bm k,\nu}^\mathrm{linear}| < \Lambda_l$),
where $\Lambda_l=\Lambda_0 \, e^{-l}$.
(b)
The initial values ($l=0$) of the RG equations 
are evaluated
 from  the cRPA analysis. The bubbles 
with thick and thin lines
 represent $\chi_{\mathrm{cRPA}}(\bm q)$ and 
 $\chi_{0\mathrm{c}}(\bm q)$, respectively.
The bare four-point vertex represents
$\Gamma^0(k_1,k_2; k_3, k_4)$,
where $k_i=(\bm k_i,\nu_i)$.
}
\label{fig:RG+cRPA}
\end{figure}

The energy band structure and 
 the FSs  obtained from $H_0$ 
 are shown in Fig.\ \ref{fig:band}.
The $\alpha$ band forms a 
holelike
FS centered at $(\pi,\pi)$ while the $\beta$ band forms an 
electronlike
 FS centered at $(0,0)$.
The RG equations are shown 
 in Fig.\ \ref{fig:RG+cRPA} (a), where 
$\chi(\bm q)$,
$R(\bm q;k_1,k_2)$, and 
$\Gamma(k_1,k_2;k_3,k_4)$
 are the
susceptibility, the three-point and four-point vertices, respectively
\cite{Bourbonnais2003,Zanchi:1998ua,Zanchi:2000ua,Halboth:2000tt}.
The scattering processes of electrons having energies less than a 
 cutoff $\Lambda_0$ are integrated within the RG scheme.
In the present analysis 
we introduce 24 patches shown in Fig.\ \ref{fig:band} (b).
As in previous works, the momenta $\bm k_i$ in 
$R$ and $\Gamma$ are projected onto 
the
Fermi surface,
shown in Fig.\ \ref{fig:band} (b).

In contrast to the conventional patch scheme \cite{Metzner:2012jv},
  we take the initial cutoff $\Lambda_0$ as a smaller value
shown in Fig.\ \ref{fig:band} (a).
Then, we treat the higher-energy contributions ($>\Lambda_0$) 
by the constrained RPA (cRPA) type method, in which 
 the high-energy interactions are included to the
infinite order for RPA-type diagrams.
In the present ``RG+cRPA scheme'',
we calculate the initial values of 
 $\Gamma$, $R$, and $\chi$
by the cRPA-type treatment,
as shown in Fig.\ \ref{fig:RG+cRPA}.
 This can be further improved
 by including the ``constrained'' VCs perturbatively
  into the initial values (cRPA+VC method).
We will show that the present RG scheme
 gives small corrections to the initial values, however, 
 provides accurate and nontrivial results for the orbital susceptibilities.

In a conventional patch RG scheme,
the higher-energy contributions are treated less accurately 
because of the projection of momenta on the Fermi surface.
In the present scheme, in contrast,
the higher-energy contributions can be accurately 
calculated perturbatively with fine $\bm k$ meshes.
Especially, this treatment is advantageous for the multiorbital model,
in which the vertices are $\bm{k}$ dependent even in the bare interactions
\cite{Metzner:2012jv}.
Thus the present scheme 
 is a natural combination of the merits of the RG (for lower energy)
and RPA-type treatment (for higher energy), 
and enables us to obtain very accurate 
susceptibilities.
This treatment is consistent with 
the Wick-ordered scheme of
the exact functional RG (fRG) formalism in
Ref.\  \cite{Metzner:2012jv}, 
so the contributions from 
the RG and cRPA are not overcounted.

In order to analyze the dominant fluctuations 
in the present two-orbital system, 
we calculate the susceptibilities by using the RG method.
The main purpose of the present Letter
 is to analyze the 
quadrupole (orbital) susceptibility at low temperatures.
In the present model, there are two irreducible quadrupole operators
\cite{Kontani:2011hf,Ohno:2013hc}:
\begin{eqnarray}
{\hat O}_{x^2-y^2}^j 
\!\! &=& \!\!\! 
\sum_\sigma
(c_{j,1,\sigma}^\dagger c_{j,1,\sigma}-c_{j,2,\sigma}^\dagger c_{j,2,\sigma})
= n_{j,1} - n_{j,2},  \quad
\label{eq:Ox2-y2}
\\
{\hat O}_{xy}^j 
\!\! &=& \!\!\!
\sum_\sigma 
(c_{j,1,\sigma}^\dagger c_{j,2,\sigma}+c_{j,2,\sigma}^\dagger c_{j,1,\sigma}),
\end{eqnarray}
where $j$ is the site index.
  The quadrupole susceptibility per spin is given by
$\chi_\gamma^\mathrm{q}(\bm q) 
= (1/2)
\int_0^{\beta} d\tau 
\langle 
{\hat O}_{\gamma}(\bm q,\tau) \, 
{\hat O}_{\gamma}(-\bm q,0) \rangle$
($\gamma=x^2-y^2$ or $xy$)
where $\tau$ is the imaginary time and $\beta = 1/(k_B T)$.
The divergence of $\chi^\mathrm{q}_{x^2-y^2}(\bm q=\bm 0)$ reflects the
emergence of the orbital nematic state ($\langle n_{xz}\rangle \neq
\langle n_{yz}\rangle$),
which is consistent with the nematic phase in Sr$_3$Ru$_2$O$_7$.
In addition,  we analyze the spin $\chi^\mathrm{s}(\bm{q})$ and charge 
 $\chi^\mathrm{c}(\bm{q})$ susceptibilities.
In order to confirm the reliability of this treatment,
we will show the results in the temperature region
   where $\chi^\mathrm{c}(\bm 0)$ remains 
nonsingular.

\begin{figure}[t]
\includegraphics[height=7.5cm]{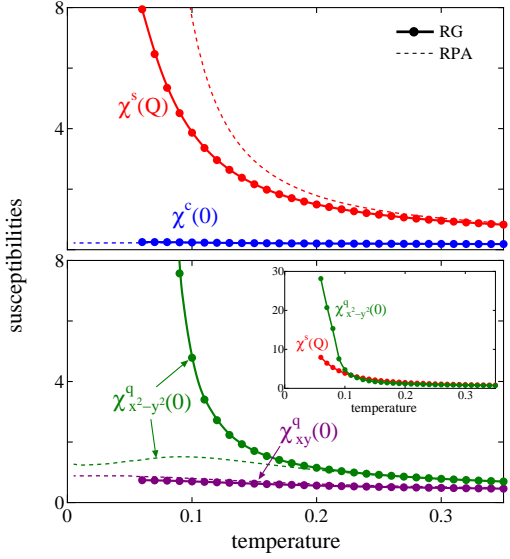}
\caption{
(color online).
 Temperature dependences of 
$\chi^\mathrm{s}(\bm Q)$, $\chi^\mathrm{c}(\bm 0)$ (upper panel), 
and
$\chi^\mathrm{q}_{x^2-y^2}(\bm 0)$, $\chi^\mathrm{q}_{xy}(\bm 0)$ 
(lower panel),
for 
$n=3.3$, $U=2.13$, and $U'/U=0.9$.
The solid (dashed) lines represent the RG (RPA) results.
In the inset, the same data of 
 $\chi^\mathrm{s}(\bm Q)$ and $\chi^\mathrm{q}_{x^2-y^2}(\bm 0)$
are plotted on a different vertical scale.
}
\label{fig:fig-chi}
\end{figure}

In the RPA without VCs \cite{Takimoto:2000ts,Ohno:2013hc}, 
$\chi^\mathrm{s\, \mathrm{RPA}}(\bm q)$  and
$\chi^\mathrm{q\, \mathrm{RPA}}_{x^2-y^2}(\bm q)$ 
are mainly enhanced by $U$ and $U'$, respectively,  and 
both of them 
 have  peak structures at 
the nesting vector $\bm q= \bm Q$ due to the nesting of FSs.
$\chi^\mathrm{s\, \mathrm{RPA}}(\bm Q)$ is always larger than
$\chi^\mathrm{q\, \mathrm{RPA}}_{x^2-y^2}(\bm Q)$ 
for the realistic parameter $U> U'$. 
We stress that $\chi^\mathrm{q\, \mathrm{RPA}}_{x^2-y^2}(\bm q=\bm 0)$ 
remains small,
meaning that the nematic
order is not realized
in the RPA.
Nonetheless,
we will show that $\chi_{x^2-y^2}^\mathrm{q}(\bm 0)$ given by
the present RG method
is strongly enhanced because of the VCs.

In order to analyze low-temperature properties accurately,
we linearize the band dispersion 
  within the cutoff scale $\Lambda_0$ and 
 change the $\bm k$ summation into the energy ($\xi$) integration
\cite{Nishine:2008wla}.
In the present numerical study, 
 we consider the case for 
 the electron filling $n=2.7$ and $3.3$,
where we choose the cutoff $\Lambda_0=0.5$ and $0.2$, respectively.

\begin{figure}[t]
\includegraphics[height=7.5cm]{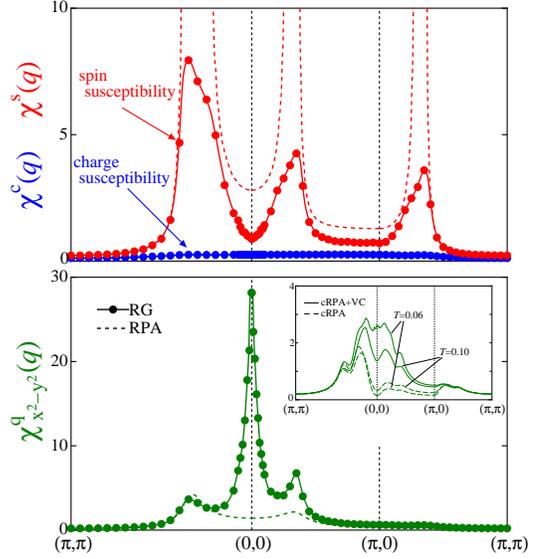}
\caption{
(color online).
Momentum dependences of 
$\chi^\mathrm{s}(\bm q)$, $\chi^\mathrm{c}(\bm q)$,
and $\chi^\mathrm{q}_{x^2-y^2}(\bm q)$
for $T=0.06$, $n=3.3$, $U=2.13$, and $U'/U=0.9$.
The solid (dashed) lines represent the RG (RPA) results.
In the inset, the momentum dependences of
 the initial values for $\chi^\mathrm{q}_{x^2-y^2}|_{l=0}$ 
 (solid lines) and the cRPA results (dashed lines) are shown.
}
\label{fig:chi-q-dep}
\end{figure}

First, we focus on the case with large filling $n=3.3$, 
where the hopping parameters are chosen as
 $(t',t_\mathrm{nn},t_\mathrm{nnn})=(0.1,0,0)$.
The best nesting vector is given by 
$\bm Q\approx (0.35\pi,0.35\pi)$.
Now, we treat the high-energy parts ($>\Lambda_0$) 
by the cRPA+VC method \cite{note:cRPA}.
The $T$ dependences of the spin, charge, and quadrupole
susceptibilities  are shown in Fig.\ \ref{fig:fig-chi}.
By making the direct comparison to the RPA results,
we can elucidate the effects of VCs.
In 
the high temperature
 ($T\gtrsim 0.3$) region,
all the susceptibilities
exhibit similar behavior to the RPA results \cite{Hirsch:1985wz}.
Even at low temperatures,
the non-singular susceptibilities
$\chi^\mathrm{c}(\bm 0)$
and $\chi^\mathrm{q}_{xy}(\bm 0)$
show the same $T$ dependences as in RPA.
These facts
   strongly indicate
 the reliability of the 
 present RG scheme.
The effect of VCs 
 suppresses $\chi^\mathrm{s}(\bm Q)$
at low temperatures.
The most striking feature of Fig.\ \ref{fig:fig-chi}
is the
critical enhancement of 
$\chi_{x^2-y^2}^\mathrm{q}(\bm 0)$  at low temperatures,
which cannot be derived from RPA.
In Fig.\ \ref{fig:chi-q-dep}, 
 the momentum dependences of 
$\chi^\mathrm{s}(\bm q)$, $\chi^\mathrm{c}(\bm q)$,
 and $\chi^\mathrm{q}_{x^2-y^2}(\bm q)$
for $T=0.06$ are shown.
Since the initial values 
 for $\chi_{x^2-y^2}^\mathrm{q}|_{l=0}$ give small contributions 
 and have weak $T$ dependences (inset of Fig.\ \ref{fig:chi-q-dep}),
the critical enhancement of $\chi_{x^2-y^2}^\mathrm{q}(\bm 0)$ 
  is achieved by the RG procedure.
We stress that the enhancement in 
$\chi_{x^2-y^2}^\mathrm{q}(\bm q)$ is restricted to the $\bm{q}\approx \bm{0}$ 
region,
indicating the emergence of the orbital nematic order.

\begin{figure}[t]
\includegraphics[width=7cm]{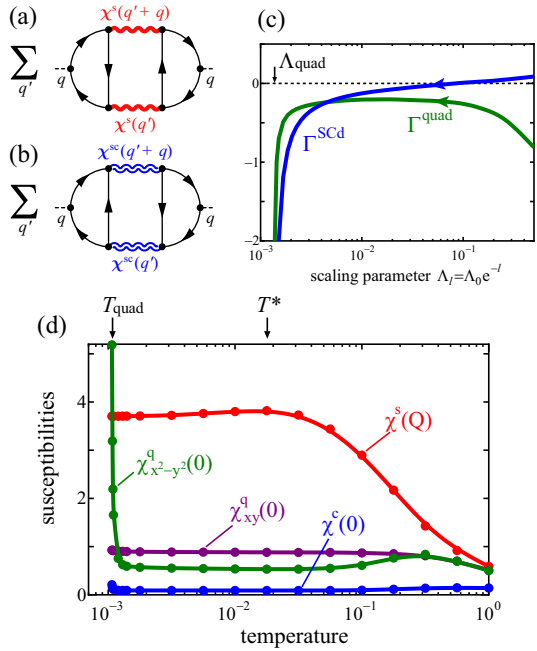}
\caption{
(color online).
(a) The magnetic AL term and (b) the superconducting AL term, where 
$\chi^\mathrm{sc}(\bm q)$ is superconducting propagator. 
They are proportional to 
$\sum_{\bm q'} \chi^z(\bm q') \, \chi^z(\bm{q}'+\bm{q})$ ($z=\mathrm{s,sc}$),
which takes  maximum  value 
 at $\bm q = \bm 0$.
(c)
The scaling flows of the effective interactions  
$\Gamma^{\mathrm{quad}}$ and $\Gamma^{\mathrm{SC}d}$
for $n=2.7$, $U=2.5$, and $U'/U=0.98$ 
at $T=10^{-10}$.
The large negative value of $\Gamma^{\mathrm{quad,SC}d}$
 gives the enhancement of the corresponding susceptibility.
(d) $T$ dependence of $\chi^\mathrm{s}(\bm Q)$,
$\chi^\mathrm{q}_{x^2-y^2}(\bm 0)$,
$\chi^\mathrm{q}_{xy}(\bm 0)$, and
$\chi^\mathrm{c}(\bm 0)$, 
for $n=2.7$, $U=2.5$, and $U'/U=0.98$.
}
\label{fig:AL}
\end{figure}

Here, we discuss the reason why 
  $\chi^\mathrm{q}_{x^2-y^2}(\bm q)$ is critically enhanced only for
 $\bm q\approx \bm 0$ in the present RG analysis.
The enhancement of the $\bm q\approx \bm 0$ mode is a strong hallmark of 
the dominant contributions of the Aslamazov-Larkin (AL) type VCs, 
since these VCs are known as the enhancement mechanisms of
$\bm q \approx \bm 0$ susceptibilities.
In Fig.\ \ref{fig:fig-chi},
both $\chi^\mathrm{q}_{x^2-y^2}(\bm 0)$  and $\chi^\mathrm{s}(\bm Q)$ 
show similar Curie-Weiss behaviors,
indicating that the enhancement of 
$ \chi^\mathrm{q}_{x^2-y^2}(\bm 0)$ originates from the spin fluctuations.
As indicated in Refs.\ \cite{Onari:2012jb,Ohno:2013hc},
the AL term due to the magnetic fluctuations 
[Fig.\ \ref{fig:AL} (a)]
gives the enhancement of $\chi^\mathrm{q}_{x^2-y^2}(\bm 0)$.
Therefore, 
the magnetic AL term of Fig.\ \ref{fig:AL} (a) is the natural origin of 
the enhancement of $\chi^\mathrm{q}_{x^2-y^2}(\bm 0)$.
As the origin of the field-induced magnetic 
quantum criticality, both the 
van Hove singularity 
\cite{Yamase:2007bh,Yamase:2009dx,Fischer:2010fo,Yamase:2010if}
and the field-suppression of quantum fluctuation \cite{Sakurazawa:2005dl}
mechanisms had been discussed previously.

Next,  we will show the importance of the 
AL term due to the superconducting fluctuations
[Fig.\ \ref{fig:AL} (b)],
which was not discussed 
in Refs. \cite{Onari:2012jb,Ohno:2013hc}.
For this purpose, we introduce the 
effective interaction
for ferro-orbital fluctuations
 $\Gamma^\mathrm{quad}$ 
and that for superconducting fluctuations
$\Gamma^{\mathrm{SC}d}$,
which give the 
 quadrupole and $d$-wave superconducting susceptibilities:
$\Gamma^\mathrm{quad} \equiv
N_\mathrm{p}^{-2} \sum_{p,p'}
[2\Gamma(p,p';p,p')-\Gamma(p,p';p',p) ]
\, \mathcal{O}_{p} \, \mathcal{O}_{p'}$ and
$\Gamma^{\mathrm{SC}d} \equiv
(1/2)
N_\mathrm{p}^{-2}  
\sum_{p,p'} \Gamma(p,\bar{p};p',\bar{p}') \,
\mathcal{D}_{p} \, \mathcal{D}_{p'}$,
where $\mathcal O_p$ is the form factor for $\gamma=x^2-y^2$ quadrupole:
$\mathcal O_p= [
  u^2_{1\nu}(\bm k(p))-   u^2_{2\nu}(\bm k(p))] |_{\nu=\nu(p)}$,
where $u_{\mu\nu}(\bm k)$ is the unitary matrix for diagonalizing 
  the kinetic term $H_0$.
Also, $\mathcal D_p$ is the $d$-wave form factor:
$\mathcal D_p=\cos k_x(p)-\cos k_y(p)$.
In the present numerical study,
we consider smaller filling ($n=2.7$),
 and put
 $(t',t_\mathrm{nn},t_\mathrm{nnn})=(0.1,0.15,0.03)$ 
\cite{Takimoto:2000ts},
where the best nesting vector is 
$\bm Q\approx (0.6\pi,0.6\pi)$.
Here, we employ the cRPA method since the VC 
for the initial values 
is not necessary
to obtain the orbital nematic fluctuations.
The scaling flows of $\Gamma^\mathrm{quad}$ and $\Gamma^{\mathrm{SC}d}$
  are shown in Fig.\ \ref{fig:AL} (c) \cite{T-flow}.
The obtained $\Gamma^{\mathrm{quad}}$ weakly depends on $l$
at high energies, while it exhibits a steep increase  in magnitude when 
$\Lambda_l = \Lambda_\mathrm{quad} (\approx 10^{-3})$.
This behavior gives the divergent behavior in 
$\chi_{x^2-y^2}^\mathrm{q}(\bm 0)$.
Moreover,
$\Gamma^{\mathrm{SC}d}$  also 
shows a steep increase in magnitude 
at $\Lambda_l\approx  \Lambda_\mathrm{quad}$:
This fact 
 means that 
the development of $\chi_{x^2-y^2}^\mathrm{q}(\bm 0)$
 is closely related to the
 superconducting fluctuations.
From the diagrammatic arguments,
we conclude that
the major contribution is given by
the superconducting AL term in Fig.\ \ref{fig:AL} (b),
which represents the coupling between the quadruple and superconducting
  fluctuations.

The $T$ dependences of the susceptibilities 
for the case of $n=2.7$
are shown in Fig.\ \ref{fig:AL} (d) \cite{note:div}.
The spin susceptibility 
$\chi^\mathrm{s}(\bm Q)$  is almost
$T$ independent below $T^* \approx 10^{-2}$
 due to the effect of nesting deviation.
Such a behavior is reminiscent of the Peierls instability in
quasi-one-dimensional
systems \cite{Duprat:2001uk,Nelisse:1999vd}.
The quadrupole susceptibility 
 $\chi_{x^2-y^2}^\mathrm{q}(\bm 0)$
shows a divergent behavior 
at $T= T_\mathrm{quad}\approx 10^{-3}$, 
which is the same energy scale of the divergent behavior
  in $\Gamma^\mathrm{quad}$ observed in Fig.\ \ref{fig:AL} (c).
In contrast,
$\chi^\mathrm{c}(\bm 0)$ and 
$\chi^\mathrm{q}_{xy}(\bm 0)$ 
show no anomaly even at $T\approx T_\mathrm{quad}$, which would ensure
the reliability of 
the singular behavior in $\chi^\mathrm{q}_{x^2-y^2}(\bm 0)$.

In summary, 
we have developed the RG+cRPA method 
for the clarification of the important effects of VCs in susceptibilities,
and 
we have confirmed 
the orbital Pomeranchuk instability
driven by magnetic and superconducting quantum criticalities.
In the large-filling case ($n=3.3$),
$\chi_{x^2-y^2}^\mathrm{q}(\bm 0)$ 
is critically enhanced by the 
AL-type VC due to the spin fluctuations,
 consistently with 
the predictions in Ref.\ \cite{Ohno:2013hc}.
In addition,
we studied the small-filling case ($n=2.7$)
and also
 found the development of $\chi_{x^2-y^2}^\mathrm{q}(\bm 0)$ 
 driven by the superconducting fluctuations.
Since the quadrupole operator $\hat O_{x^2-y^2}$ represents 
the order parameter of 
  the orbital ordering [Eq.\ (\ref{eq:Ox2-y2})],
both mechanisms would contribute to 
the nematic phase in  Sr$_3$Ru$_2$O$_7$.
The present mechanisms of
the orbital nematic phase would be realized 
  in various multiorbital systems.

The authors 
are thankful for
fruitful discussions with 
W.\ Metzner, C.\ Honerkamp, 
M.\ Sigrist, C.\ Bourbonnais, A.-M.S.\ Tremblay, H.\ Yamase, 
Y.\ Yamakawa, and  D.S.\ Hirashima.
This work was supported by a Grant-in-Aid for Scientific Research from 
the Ministry of Education, Culture, Sports, Science, and Technology, Japan.

\end{document}


\title{
Supplemental Material \\ 
Initial function due to cRPA+VC method
}

\author{Masahisa Tsuchiizu}
\affiliation{Department of Physics, Nagoya University, Nagoya 464-8602, Japan}

\author{Yusuke Ohno}
\affiliation{Department of Physics, Nagoya University, Nagoya 464-8602, Japan}

\author{Seiichiro Onari}
\affiliation{Department of Applied Physics, Nagoya University, Nagoya 464-8603, Japan}

\author{Hiroshi Kontani}
\affiliation{Department of Physics, Nagoya University, Nagoya 464-8602, Japan}

\maketitle

In the present RG method, we introduce the cutoff energy $\Lambda_0$
according to the functional RG theory \cite{Metzner:2012jv}.
We set $\Lambda_0$ so as that
the topology of the constant energy surfaces 
$\xi_{\bm{k},\nu}=\varepsilon$ 
($\xi_{\bm{k},\nu}$ being the $\nu$-th band dispersion)
for $|\varepsilon|<\Lambda_0$ is equivalent to the Fermi surface.
In the present work, the scattering processes under the cutoff 
$\Lambda_0$ are calculated by solving the one-loop RG equations,
by dividing the Fermi surfaces into $N_\mathrm{p}$ patches.
The four-point vertex with higher-energy contributions,
$\hat\Gamma^{\rm H}
(\bm{p}\sigma,\bm{p}'\sigma';\bm{k}\rho,\bm{k}'\rho')$, 
is incorporated as the initial values
of the RG differential equations,
where $\bm{k}$ and $\bm{p}$ ($\sigma$ and $\rho$)
 are momentum (spin) indices.
According to the functional RG theory \cite{Metzner:2012jv},
$\hat\Gamma^{\rm H}$ is constructed by 
the Coulomb interaction ($U,U',J$) in addition to the 
higher and lower energy Green functions,
$G_\nu^{\rm H}(\bm{k})\equiv G_\nu(\bm{k})
 \, \Theta(|\xi_{\bm{k},\nu}|-\Lambda_0)$ and
$G_\nu^{\rm L}(\bm{k})\equiv G_\nu(\bm{k})\, 
\Theta(\Lambda_0-|\xi_{\bm{k},\nu}|)$ respectively,
under the constraint that $\hat\Gamma^\mathrm{H}\!$ 
is irreducible with respect to the product of $\!G_\nu^\mathrm{L}\!$.

By virtue of introducing the cutoff energy $\Lambda_0$,
we can perform the RG calculation with high accuracy 
since the topology of the constrained energy surfaces is unchanged 
in the lower energy regime.
We also calculate the higher-energy contributions
by using the perturbative methods.
In the present study, we apply (i) the constrained RPA (cRPA) method,
and also apply (ii) the cRPA method with higher-order vertex corrections.

Hereafter, we derive the expression of $\Gamma^{\mathrm{H}}$ 
on the $d$-orbital basis, 
which should be converted to the 
patch index  to use as the initial values of the RG equations.
The spin and charge susceptibilities in the cRPA are given 
by the following $(N_{\rm orb}^2\times N_{\rm orb}^2)$ matrix form
($N_{\rm orb}$ is the number of orbital degrees of freedom):
\begin{eqnarray}
\hat \chi^\mathrm{s(c)}_\mathrm{cRPA}(q)
=\frac{{\hat \chi}_{0\mathrm{c}}(q)}
{1-{\hat \Gamma}^\mathrm{s(c)}{\hat \chi}_{0\mathrm{c}}(q)},
\end{eqnarray}
where ${\hat \Gamma}^\mathrm{s(c)}$ is the bare interaction for the 
spin (charge) channel given in Ref. \onlinecite{Onari-SCVC}, 
and $q=(\bm{q},2\pi lT )$.
The constrained bare susceptibility 
${\hat \chi}_{0\mathrm{c}}(\bm{q})$ is given by
\begin{eqnarray}
\left[{\hat \chi}_{0\mathrm{c}}(q)\right]_{l,l';m,m'}
&=&-T\sum_{\bm{k}} G_{l,m}(k+q) G_{m',l'}(k)
 \nonumber \\
& & +T\sum_{\bm{k}} G_{l,m}^{\rm L}(k+q) G_{m',l'}^{\rm L}(k),
\end{eqnarray}
where 
$l,l',m,m'$
are the orbital indices, 
$k=(\bm{k},\pi(2n+1)$ $ \pi T)$,
and 
$G_{l,m}^{(\mathrm{L})}(\bm{k})$ 
is the Green function on the orbital basis
given by the unitary transformation from $G_{\nu}^{(\mathrm{L})}(\bm{k})$.

In method (i), we approximate 
$\!\Gamma_{l,l';m,m'}^\mathrm{H}
  \!(\bm{p}\sigma,\bm{p}' \! \sigma'\! ; \!\bm{k}\rho,\bm{k}'\! \rho')$ 
in the $(N_{\rm orb}^2\times N_{\rm orb}^2)$ matrix form as
\begin{eqnarray}
&&{\hat \Gamma}^{\rm H}(\bm{p}\sigma,\bm{p}'\sigma';\bm{k}\rho,\bm{k}'\rho')
= {\hat \Gamma}^{0}_{\sigma,\sigma';\rho,\rho'}
\phantom{\frac{ZZZ}{\frac{ZZZ}{ZZZ}}}
\nonumber \\
&&\hspace*{1cm}
+{\hat V}_{\sigma,\sigma';\rho,\rho'}(\bm{p-p}')
-{\hat V}'_{\sigma,\sigma';\rho,\rho'}(\bm{p-k}),
\qquad
 \label{eqn:Gh}
\end{eqnarray}
where ${\hat \Gamma}^{0}$ is the antisymmetrized bare
Coulomb interaction with spin indices ($\sigma,\sigma',\rho,\rho'$).
Considering the SU(2) symmetry of the spin space, it is expressed as
\begin{equation}
{\hat \Gamma}^{0}_{\sigma,\sigma';\rho,\rho'}
\equiv \, {\hat \Gamma}^\mathrm{c} \, \delta_{\sigma,\sigma'}\delta_{\rho',\rho}
+{\hat \Gamma}^\mathrm{s} \, {\bm \sigma}_{\sigma,\sigma'}\cdot {\bm \sigma}_{\rho',\rho},
\end{equation}
where ${\bm \sigma}$ is the Pauli matrix vector.
Also, ${\hat V}$ is given as
\begin{eqnarray}
{\hat V}_{\sigma,\sigma';\rho,\rho'}(\bm{q})=
{\hat V}^\mathrm{c}(\bm{q},0) \, \delta_{\sigma,\sigma'}\delta_{\rho',\rho}
+{\hat V}^\mathrm{s}(\bm{q},0) \, {\bm \sigma}_{\sigma,\sigma'}\cdot {\bm \sigma}_{\rho',\rho},
\label{eq:V}
\end{eqnarray}
where 
${\hat V}^\mathrm{s(c)}(q)={\hat \Gamma}^\mathrm{s(c)} \,
{\hat \chi}_\mathrm{cRPA}^\mathrm{s(c)}(q) \, {\hat \Gamma}^\mathrm{s(c)}$.
In addition,
$[{\hat V}'_{\sigma,\sigma';\rho,\rho'}(\bm{q})]_{l,l';m,m'}
\equiv 
[{\hat V}_{\sigma,\rho;\sigma',\rho'}(\bm{q})]_{l,m;l',m'}$.

In method (i), the initial value in Eq.\ (\ref{eqn:Gh})
is underestimated for $\bm p = \bm p'$ and $\bm k= \bm k'$
 because of the reason
${\hat \chi}_\mathrm{cRPA}^\mathrm{s(c)}(\bm{0},0)\approx0$ for $T\ll\Lambda_0$.
This relation in the cRPA is unfavorable for the development of 
the susceptibilities at $\bm{q}=\bm{0}$,
by solving the RG equations.
However, this relation is an artifact of the cRPA,
and it will not be satisfied in general.
In method (ii), therefore, we improve the initial value $\Gamma^{\rm H}$ 
in Eq.\ (\ref{eqn:Gh}), by introducing the vertex correction
into the cRPA as 
\begin{eqnarray}
{\hat \chi}^\mathrm{s(c)}_\mathrm{cRPA+VC}(q)
=\frac{{\hat \chi}_{0\mathrm{c}}(q)+{\hat X}^\mathrm{s(c)}(q)}
{1-{\hat \Gamma}^\mathrm{s(c)}
  [{\hat \chi}_{0\mathrm{c}}(q)+{\hat X}^\mathrm{s(c)}(q)]},
\label{eq:cRPA+VC}
\end{eqnarray}
where 
${\hat X}^\mathrm{s(c)}(q)$ is the vertex correction (VC)
due to higher-energy contribution.
Then, the initial function is improved by replacing 
${\hat \chi}_\mathrm{cRPA}(q)$ 
with 
${\hat \chi}_\mathrm{cRPA+VC}(q)$
in Eq.\ (\ref{eqn:Gh}).
Here, we calculate the VC
up to the second-order diagrams with respect to 
${\hat \chi}_\mathrm{cRPA}(q)$:
The first-order term is called the Maki-Thompson (MT) term,
and the second-order term is called the 
Aslamazov-Larkin (AL) term \cite{Onari-SCVC}.
We find that the MT term is negligibly small, and only the 
AL term for the charge channel is quantitatively important.
It is given by
\begin{widetext}

\vspace*{-.8cm}
\begin{equation}
X_{l,l';m,m'}^\mathrm{c}(q)
=\frac{T}2\sum_{k}\sum_{a\sim h}
\Lambda_{ll',ab,ef}(q;k)
\Bigl[
{V}_{ab,cd}^\mathrm{c}(k+q) \, {V}_{ef,gh}^\mathrm{c} (-k)
 +3{V}_{ab,cd}^\mathrm{s}(k+q) \, {V}_{ef,gh}^\mathrm{s}(-k) 
\Bigr]
\Lambda_{mm',cd,gh}'(q;k) .
 \label{eqn:ALexample}
\end{equation}

\noindent
Here, ${\hat \Lambda}(q;k)$
 is the three-point vertex due to 
higher-energy contribution, given as
\begin{eqnarray}
&&\Lambda_{mm',cd,gh}(q;k)
= 
\sum_{A,B,C}
T\sum_{p} G_{m,c}^A (p+q) \, 
G_{d,g}^B(p-k) \, 
G_{h,m'}^C(p),
 \label{eqn:threeP}
 \end{eqnarray}
\end{widetext}
where
$p=(\bm{p},\pi(2n'+1)T)$, and 
 $\Lambda_{mm',cd,gh}'(q;k)\equiv
\Lambda_{ch,mg,dm'}(q;k)+\Lambda_{gd,mc,hm'}(q;-k-q)$.
The summations with respect to  $A,B,C$ are taken as
$(A,B,C)=(\mathrm{H,H,H})$, $(\mathrm{L,H,H})$, $(\mathrm{H,L,H})$, 
$(\mathrm{H,H,L})$.
Here, $\hat{X}^\mathrm{c}(q)$
 is irreducible with respect to 
the product of $G^\mathrm{L}_{l,m}(k)$.

Within the cRPA, the initial interaction $V(\bm{q})$ 
in Eq.\ (\ref{eq:V}) is small for 
the forward scattering ($\bm{q}=\bm{0}$) at low temperatures since
$\chi_{0\mathrm{c}}(\bm{q},\omega=0)|_{\bm{q=0}}  
=\sum_{\bm{k}} (-df/d\epsilon)_{\epsilon=\xi_{\bm{k}} }
\theta(|\xi_{\bm{k}}|-\Lambda) \approx 0$
at $T\ll \Lambda$.
However, 
the VC $X(\bm{q})$
 in Eq.\ (\ref{eqn:ALexample}) is finite
at  $T\ll \Lambda$, since the constrained three-point vertex in Eq.\
(\ref{eqn:threeP})
 is finite
even for $\bm{q}\approx \bm{0}$,
 which can be proved analytically as well as numerically.
Therefore,
 the VC gives a finite contribution to 
  ${\hat \chi}^\mathrm{c}_\mathrm{cRPA+VC}(\bm{0})$,
as seen in Eq.\ (\ref{eq:cRPA+VC}), and 
the smallness of ${\hat \chi}^\mathrm{c}_\mathrm{cRPA}(\bm{0})$
 is an artifact of the cRPA due to the 
  absence of the VC.

Diagrammatically, ${\hat X}^\mathrm{s(c)}(q)$ is the second-order term
with respect to 
 ${\hat \chi}_\mathrm{cRPA}(q)$
composed of 
multiple $G= (G^\mathrm{H},G^\mathrm{L})$:
The latter constructs the three-point vertex in Eq.\ (\ref{eqn:threeP}).
In calculating Eq.\ (\ref{eqn:threeP}),
 we introduce another cutoff energy $\Lambda_1(> \Lambda_0)$
and put $G_\nu(k)=0$ for $|\xi_{\bm{k},\nu}|>\Lambda_1$ in the band-diagonal basis:
Although the VC is underestimated by introducing $\Lambda_1$,
we can drop the almost momentum-independent contribution to the VC,
which is unimportant for the RG result.
In the present study, we put $\Lambda_1=0.6\ (=3\Lambda_0)$:
the obtained ${\hat X}^\mathrm{s(c)}(q)$ is very similar
for $\Lambda_1=0.4\sim1.0$.

In the present method,
the contributions from 
the RG and the constrained perturbation  
are not overcounted,
since $\hat\Gamma^\mathrm{H}$ in this study is constructed consistently with 
the Wick-ordered scheme of the exact functional RG formalism
\cite{Metzner:2012jv}.
In addition,
${\hat X}^\mathrm{s(c)}(\bm{q})\ll {\hat \chi}_{0\mathrm{c}}(\bm{q})$ 
except for 
$q_\mu\approx 0$ ($\mu=x,y$),
 and the smallness of ${\hat \chi}_{0\mathrm{c}}(\bm{q})$
at $q_\mu\approx 0$ is an artifact of the approximation,
as shown in Fig.\ 4 (b) in the main text.
This fact means that
the perturbative treatment with respect to 
${\hat \chi}_\mathrm{\rm cRPA}(\bm{q})$ given by the present study is justified.
The large nematic susceptibility given by the present work 
originates from the development of the low-energy four-point vertex
given by the RG equations.
By virtue of the present RG+cRPA type approximation,
we obtain beautiful temperature- and momentum-dependences of 
the susceptibilities numerically.